# PATTERN LANGUAGE FOR GOOD OLD FUTURE FROM JAPANESE CULTURE

Megumi Kadotani
Aya Matsumoto
Takafumi Shibuya
Younjae Lee
Saori Watanabe
Takashi Iba

Faculty of Policy Management, Keio University
5322 Endo, Fujisawa,
Kanagawa, Zip 252-0882,Japan
e-mail: s11214mk@sfc.keio.ac.jp

**ABSTRACT**

Having developed greatly over millennium under its culture, the ancient buildings and old town atmospheres maintain a quality of comfort. However, people only appreciate the "good old" quality and do not think further about the rational reasons why they feel comfort in it. This keeps them from creating their own things and models with good old quality, relying on the imported western thinking and methods as a result of modernization. Since people are now struggling under the imbalanced and complex society, we believe that support is needed for generating things and frameworks with good old quality in modern time situations and scenes.

Therefore, we propose a language, which verbalized the process of generating things and systems with such quality under creative activities. Based on Christopher Alexander's pattern language of describing good design practices, we have created a Pattern Language for good old future from Japanese culture. In order to create things and frameworks with a quality of comfort in collaboration, a common language is needed among people in Japan. Using this pattern language, people will be able to create their own comfortable society where frameworks and things with good old quality are widespread. Additionally, by extracting the good old quality from old cultural backgrounds, it will be possible to build on a society with its fully cultivated culture alongside a firm basis.

**INTRODUCTION**

Nowadays, people visit ancient buildings and towns as sightseeing places where they feel and enjoy the atmosphere and emotional senses. We unconsciously feel comfort in and appreciate such mood called good old quality, which has developed over a period of time. It is a quality, which we feel comfort in while usually visiting traditional and historical sites, in thought of the past.

However, people go no further than sensing the good old quality, leaving behind the deep understandings about the reasons why we feel comfortable.

Though our ancestors has accumulated the comfort in traditional culture, if we do not understand the reasons why we feel pleasure in it, finding ways to create a future with comfort would be difficult.

More importantly, we are most likely to design a future with comfort not only by finding and understanding the good old quality from the past, but also by setting up creations using a suitable method. This will help pass down the comfort to the next generation and set precedents of good old periods and society to reflect on.

Naming such comfort, good old quality, we eagerly aim to support people in both discovering it and making it.

To accomplish our goal of helping people to discover and create a good old quality, we used the format of pattern language as an effective description.

**PATTERN LANGUAGE**

Architect C. Alexander pursued the quality of "good old" in buildings and has presented pattern language, a methodology of describing good design practices. He extracted a common form out of good old buildings and called it a "pattern." After gathering them as one language, C. Alexander presented pattern language, which is verbalized and described in a particular format. Each pattern is written in the order of Context, Problem, Solution, and Consequence, explaining the types of problems and its occasion, possible actions, and finally, the result. Following the Alexandrian format, context, problem, and solution, help people realize, as context, solution, and consequence, influence people to design. Therefore this inspired people both to discover and create a good old quality.

Moreover, by stating the solution, problem, and consequence, this format clarifies what method makes us feel pleasant. This rational reason will engage more people to interpret the good old quality, regardless of stereotypes and bias.

However, it is meaningless to claim that past Japanese method holds a good old quality because qualities are unmeasurable. With that said, we believe that we can judge with an objective and modern perspective, whether the quality exists or not.

## PATTERN LANGUAGE FOR FUTURE FROM JAPANESE CULTURE

### Making Process
In this section, we would like to introduce the process of making such pattern language. At the Iba Laboratory of Keio University Shonan Fujisawa Campus in Japan, research is being conducted on designing a system for generating a creative society using pattern language. This spring, 6 students started "good old future protect," a project to verbalize the process of creating the quality of "good old".

*Mining*
We have navigated each member to his/her interested fields: working style, discrimination over poverty, education, urban design, aesthetic, gourmet, leisure, etc. and assigned related works, literature, and references from Japanese culture. Then, we mined examples of our ancestors' rich lives, which are being forgotten over time, and listed each on sticky notes.

*KJ Method*
We have pulled out up to 250 examples and grouped them by using a convergent thinking called the KJ method (Kawakita, 1976). This technique is used to arrange the complex collection of examples and cases according to their similarity. Focusing on points that consist of good old quality, we have discussed the degree of similarity and organized the sticky notes accordingly into 44 groups. Addressing each group with a name, we have identified 12 pattern models and completed them as 12 patterns, which are stated below.

>12 Patterns
-COMPARISON BY SHAPE
-THE POWER OF AMBIGUITY
-TRANSIENT DREAM
-LIVE BY WATERS
-REVERSE VOID AND FULL
-INNERMOST AND EXTRAORDINARY
-THE SENSE OF AWE
-FIT IN AND BREAK FORMS
-LET IT BE
-MICROCOSMOS REPRODUCTION
-COMPASSION FOR SURROUNDINGS
-PATCH THE IMPERFECTION

Documenting the results of our discussion using KJ method's format, we present the details of 3 patterns in the next chapter.

However, since these are still work in progress, we would like to address our future plans in completing our tasks.

*Field Work*
Since we live in modern time and have obtained elements from references, it is necessary to go out to the field and understand the hints lying in the quality of "good old." In order to elaborate on the content with persuasiveness, there is a need to undergo such activities.

Thus, we are planning to conduct field works out to places where we can experience the good old atmosphere. For example, we have bucket-listed not only the historical sightseeing places such as Kamakura, Kyoto, and Asakusa but also the outskirts of Tokyo, rural woodlands, and local farms.
As we plan above, exploring the sights will help us confirm whether we can acknowledge the good old quality.

*Writing & Revising*
In result of the KJ method, when we have confirmed that each group has the quality of "good old" through some field works, we will document the pattern models based on the Alexandrian format. Then, we will hold a writer's workshop after completing to write each pattern. This workshop is a method to review, evaluate, and improve pattern descriptions. The participants read through the patterns and the author will remain silent and does not speak. During the workshop the participants can discuss with objective views about the purposes of each pattern.

This method is effective in enhancing the quality of patterns with a concrete message and in engaging people to improve each pattern together.

**Prototypes**
Below, we will introduce 3 out of all patterns that we has made.

No.1 MICROCOSMOS REPRODUCTION
No.2 INNERMOST AND EXTRAORDINARY
No.3 REVERSE VOID AND FULL

No.1

## MICROCOSMOS REPRODUCTION

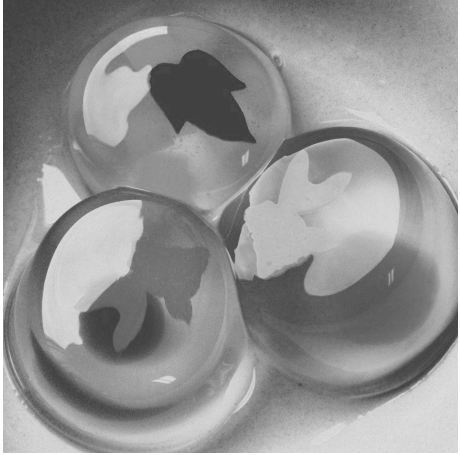

<Scene>
"Bonsai," originated in and imported from China, had developed as part of the important Japanese culture to cherish the sceneries of nature through plant cultivation in gardens. After the miniature tree is transferred to a low-sided pot and is gently taken good care of, it allows the viewers to imagine it as a full-size tree with a compact form of aesthetics. The unique practice of looking further into the "unseen" aesthetics of Bonsai shows the tacit interaction with miniature trees. The purposes of bonsai are primarily contemplation and self-realization on a grand scale through continuous dialogues with the traditional art form. Additionally, Bonsai transforms its theme each season, placed at Japanese-style restaurants to demonstrate the seasonal aspects of nature. Such display offered an opportunity for the viewers to talk about the theme and added a cultural topic in communication.

In modern days, small figures of "anime" characters are presented as specimen of the unreal world. Another example would be miniature trains, which exemplifies the big and long trains to enjoy at home. As explained, the specimen inspires us to conceptualize a reviving and vital image, capturing the details of the object and converting it into a smaller size as duplication. Also, well-known Japanese fresh confections in Kyoto are skillfully made to vastly express note a seasonal reminder in spite of being small. In making Japanese confectionary, diverse and profound art techniques are used, such as representing the season by color and highlighting contrast to symbolize complex nature.

<Context>
You come across a mesmerizing scene.

<Problem>
You are trying to keep the memory that you have encountered in real life, but you forget as time elapses.

<Solution>
**Exemplify and compress the grand scenery into a smaller size as a compact universe in details.**

<Consequence>
You can immerse yourself into the grand scenery through an art form in a small scale. Such form of Japanese aesthetic will cover all aspects of the scenery in every detail. This will allow the viewers to witness the vitality and lively objects as if they are moving. The grandness of a microcosm and tradition will originate under the co-existence of complexity and authenticity.

No.2

## INNERMOST AND EXTRAORDINARY

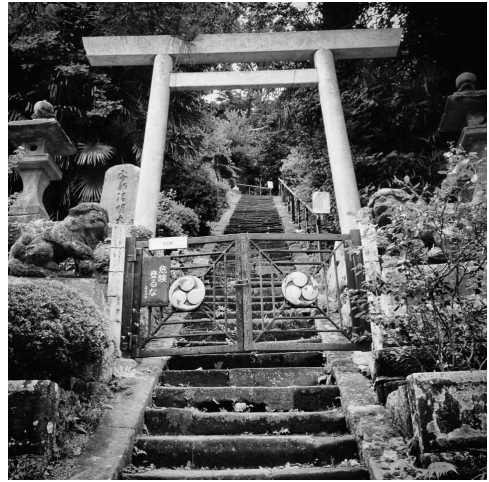

<Scene>
Like the inner shrine of Mount Koya, the residence of the Gods has been placed behind the main hall of the temple or deep in the woods from ancient times. The path to the gods' residence runs through the shrines in the order of *"Shimosya," "Nakasya," "Kamisya," and finally "Okusya." Thus, it is as if the path is guiding us towards an uncommon world from daily life.

Traditionally constructed houses are also made with spacing of depth, drawing us on a long way from the entrance to the living room, the hallway to the Japanese parlor, then to the bedroom. Therefore, this distance emphasizes "depth" in any space and tracing such area was considered good.

Furthermore, Basho Matsuo, writer of "haikai" poems, expresses aesthetic and dignity, in one of his works; "Through the curtains of the deep inner room, I catch a glimpse of the northern ume." As seen, people have been valuing the concept of depth in time and space.

*Japanese shrines are named depending on the height they are placed.
E.g. Shimosya (lowland area,) Nakasya (middle area,) Kamisya (upper part of land,) and Okusya (Deep from the main hall.)

<Context>
There are places to worship god and to offer prayers.

<Problem>
Since you want to keep the place a private space but other people also think alike, a crowd will occur.

<Solution>
**Create an inmost space and time to deeply delve into the objects.**

<Consequence>
Because such "depth" is unidentifiable, it arouses both mental states: curiosity that inspires people to further investigate and the fear to know the answer. Once in a nervous situation, people who strive for further engagement tend to surpass those who don't challenge themselves. Additionally, you can prepare for greeting god with a calm mindset, feeling your thoughts and emotions circulating inside yourself due to the pressure to proceed.

No.3

**REVERSE VOID AND FULL**

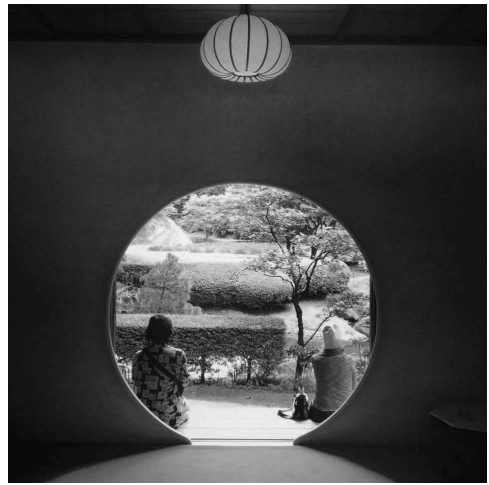

<Scene>
A Japanese term, "Mono no aware" implies the pathos aesthetics in the transience of things and bitter sadness of their leave. Since Japanese people already knew that all creations never remain, and accepted such reasonable theory, they valued abiding it. Along time ago, traditional Japanese houses had functionless and hallow spaces called "Utsu." In that place, low-dinning tables were placed for dinner, a knitting box was set up to do needle work during lunch, and Futon was spread to sleep at night. Though having limited amount of space, our ancestors took advantage of it in a flexible manner using a variation of tools and flow of the day. As explained, the ways of adding meanings and function to space were exemplified in daily lives.

<Context>
You need to remain at one and place.

<Problem>
Since you need to adjust to various changes, you can need to keep coming up with new ideas every time.

<Solution>
**Purposely leave space in the room adaptable to change.**

<Consequence>
A space with hallow called "Utsu," can also be a platform of "Utsuroi," transforming into a place for change. In other words, people would be able to become more flexible by overcoming sudden changes rather than blocking them away in a negative attitude. This will also enable them to remember excitements of one-and-only moments, reminding them that nothing stays the same. Moreover, with this in mind, a view of life as something transient and empty is generated.

**<u>CONCLUSION</u>**

There are two advantages of describing and sharing pattern language. First, by documenting the experiences and precedents, a creative problem-solving method can be inherited to the future generations.

Another would be by suggesting a common vocabulary and language on designing, people can extend their communication about complex relations and qualities unrevealed while having a common knowledge basis.

As you can see from our Pattern Language for good old future from Japanese culture, it is possible to extract and write up a language from the good old qualities of Japanese traditions. Therefore, we expect this pattern language to be a common language among people who design things and frameworks to create a good old society. This will trigger an inspiring collaboration and discussion that generates things with good old qualities.

Additionally, people's creations will vary depending on the existence of such quality, which will guide people as a new indicator to repair and improve their creations.

**FUTURE WORK**

As this pattern language for good old future from Japanese culture is currently in progress, we aim to complete it by November 2013.

After finishing all the process, we will share the making process with public as an open source, engaging people who wish to collaborate with us. For example, we would like to improve or multiply our patterns by interviewing masters of traditional Japanese dance or drama and craftsmen who are in need to transfer the knowledge onto future generations. In addition, we hope to attract art students or youth interested in history and culture to advertise the patterns by designing workshops through hands-on experiences.

Lastly, we hope people can collaborate and generate things and frameworks with good old qualities using this common language.

With that said, people will not only value the pattern language we suggested but also continuously create a pattern language adjustable and suitable to its period. Also, people can continue generating ideas that have good old qualities and keep updating the language appropriate to its era even 20 or 30 years later.

Acquiring the method to create good old qualities applicable to all areas and era, we believe that there could be a good old future, full of things and frameworks with a good old quality.

**ACKNOWLEDGEMENT**

In finishing this paper, we would like to appreciate the support of Tomoki Furukawazono who gave us much advice and Sumire Nakamura who translated our paper into English. Also, we would like to extend our deep gratitude to Kazuya Ohara, Daiki Muramatsu, Shouta Seshimo, Ko Matsuduka who also spent time revising our paper. In addition, we are very grateful to all members in Iba laboratory at Keio University for giving us many comments and support.

**REFERENCES**

Gloor, P. (2010) Coolfarming: Turn Your Great Idea into the Next Big Thing
Amacom Books
Alexander, C. (2003) The Phenomenon of Life: The Nature of Order, Book 4: An Essay of the Art of Building and the Nature of the Universe
Routledge
Alexander, C. (2003), The Process of Creating Life: The Nature of Order, Book 2: An Essay of the Art of Building and the Nature of the Universe
Routledge
Alexander, C. (2003), A Vision of a Living World: The Nature of Order, Book 3: An Essay of the Art of Building and the Nature of the Universe
Routledge
Alexander, C. (2003), The Luminous Ground: The Nature of Order, Book 4: An Essay of the Art of Building and the Nature of the Universe
Routledge
Alexander, C., Neis, H., and Alexander, M., M. (2012), The Battle for the Life and Beauty of the Earth: A Struggle Between Two World-Systems, Oxford University Press
Alexander, C., Ishikawa, S., Silverstein, M., Jacobson, M., Fiksdahl-King, I., and Angel, S. (1977), A Pattern Language: Towns, Buildings, Construction, Oxford University Press
Alexander, C. (1979), The Timeless Way of Building, Oxford University Press
Iba, T. (2010), "An Autopoietic Systems Theory for Creativity", Procedia - Social and Behavioral Sciences, Vol.2, Issue 4, pp.6610-6625
Charles, H., Paul, W. and Arthur,W., B. (1998), Collected Papers of Charles Sanders Peirce, Thoemmes Press
Kawakita, J. (1976), The Original KJ Method, Tyukou-shinsho